\documentclass[runningheads]{llncs}
\usepackage{orcidlink}
\usepackage{hyperref}
\usepackage[T1]{fontenc}
\usepackage{amsmath,graphicx,verbatim}
\newcommand{\etal}{\textit{et~al.~}}

\begin{document}
\title{Expert-like Bone Ultrasound Segmentation through Expert-in-the-loop Mask-conditioned Progressive Learning}
\titlerunning{ExiL Progressive Learning for Image Segmentation}
%

\author{Arash Tavangar\inst{1} \and
Larissa K. Chiu\inst{1}\orcidlink{0000-0002-9370-7157} \and
Hamidreza Khodashenas\inst{2} \and
Gregory K. Berry\inst{3}\and
Amir Hooshiar\inst{1,2 *}\orcidlink{0000-0002-9036-5331}}

\index{Tavangar, Arash}
\index{Chiu, Larissa K.}
\index{Khodashenas, Hamidreza}
\index{Berry, Gregory K.}
\index{Hooshiar, Amir} 

\authorrunning{Tavangar et al.}

\institute{
Department of Surgery, McGill University, Montreal, QC, Canada\\
\and
Research Institute of the McGill University Health Centre, Montreal, QC, Canada\\
\and
Division of Orthopedic Surgery, Department of Surgery, McGill University, Montreal, QC, Canada\\
*corresponding author: \email{amir.hooshiar@mcgill.ca}\\
}

\maketitle              
\begin{abstract}
Manual annotation remains a major bottleneck in ultrasound (US) bone segmentation, where experts typically iteratively refine rough brush masks rather than delineating precise contours in a single pass. We present ExiL, a mask-conditioned progressive learning framework that models annotation as a structured refinement trajectory. ExiL combines a synthetic expert-like brush simulator based on signed distance fields with a lightweight 7.8M-parameter U-Net that learns to complete and refine imperfect masks from US images. During deployment, an expert mode updates the model directly from accepted refinements, enabling continual adaptation to expert behavior. Evaluated using UltraBones100k cadaver data for quantitative segmentation and a prospective volunteer dataset for annotation-efficiency analysis, ExiL reduced single-expert average annotation time from 60 to 20 seconds per frame (66.7\%) and improved mean Dice by approximately 0.045 over non-progressive training, while achieving 0.87 Dice and 2.7 px boundary error in the best trajectory-aware setting. With 10--50 ms inference, ExiL enables real-time, self-improving annotation for US-guided orthopedic workflows in practical clinical labeling.

\end{abstract}


\section{Introduction}
\begin{figure}[t]
    \centering
    \includegraphics[scale=0.35]{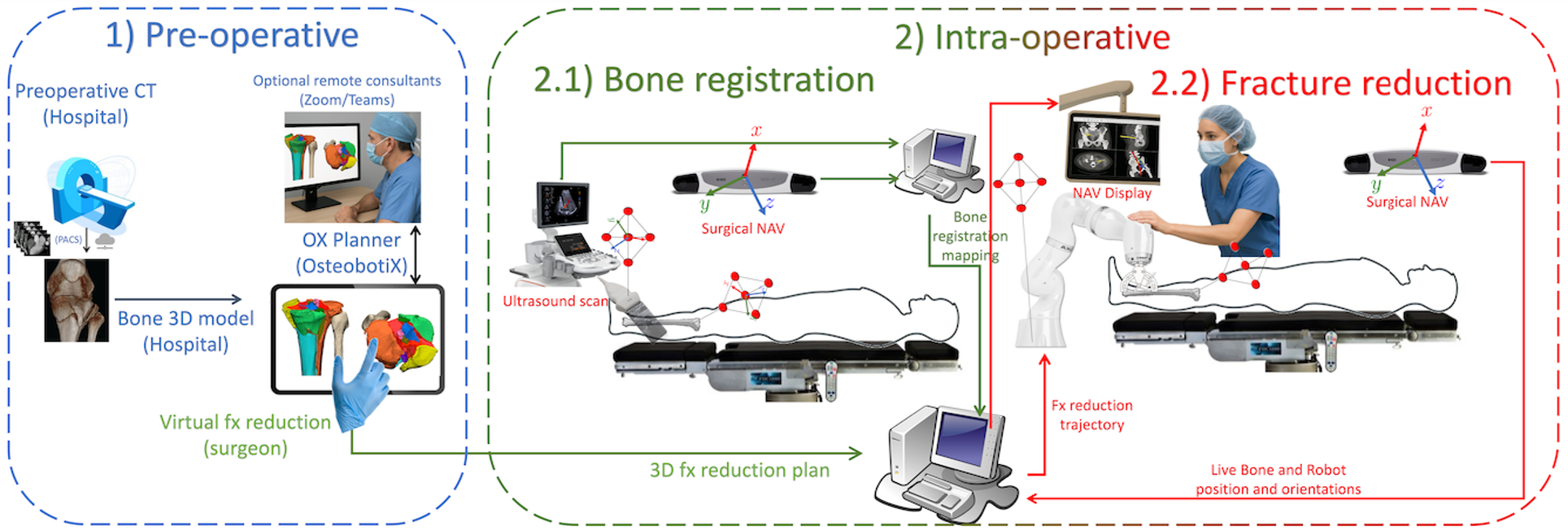}
    \caption{A representative operational workflow of image-guided surgical robotic platform using US image registration for common orthopedic procedures.}
    \label{fig:workflow}
\end{figure}

\paragraph{Clinical Need and Workflow}: Ultrasound imaging (US) offers real-time imaging at low cost and without ionizing radiation, making it particularly attractive for intraoperative and point-of-care applications \cite{endara-minaComparativeUseUltrasound2023}. Recently, studies have suggested US-based bone registration with preoperative and intraoperative workflows, as depicted in Fig. \ref{fig:workflow}. To this, bone surface segmentation in US images plays a crucial role  \cite{hacihaliloglu2017review,noble2006survey}. 
Manual segmentation by expert clinicians remains the gold standard for delineating anatomical structures in medical imaging \cite{maSegmentAnythingMedical2024}. However, it is time-intensive, especially in US, where speckle noise, low contrast, and acoustic shadowing hinder precise boundary definition \cite{hacihaliloglu2017review,alsinanAutomaticSegmentationBone2019}. Even for well-defined bone interfaces, inter- and intra-observer studies report sub-millimetre variability \cite{villaFCNbasedApproachAutomatic2018,dutoitAutomaticFemoralArticular2022}. Thus, while reliable, manual segmentation is labour-intensive and difficult to scale. Most existing methods still frame segmentation as a static prediction task and depend on high-quality labels plus direct post-processing to complete masks \cite{wang2018interactive,alsinanBoneShadowSegmentation2020}. Although robustness to noisy labels has been studied \cite{karimi2020noisy,yao2023spatialcorrection}, these methods mainly improve training stability rather than support experts during annotation. Interactive methods use clicks or scribbles, but they typically treat each interaction independently and do not learn the expert’s evolving annotation behavior \cite{buddSurveyActiveLearning2021,xu2025you,wong2024scribbleprompt,wong2025multiverseg,mikhailov2024deep}.
\paragraph{Related Works}: Label noise in medical segmentation is often spatially structured rather than independent. Yao \etal addressed boundary-correlated noise with spatial correction to improve training robustness \cite{yao2023spatialcorrection}, while Chen \etal and Fan \etal proposed refinement networks based on synthetic error augmentation and diffusion-based correction, respectively \cite{chen2025lrn,fan2025vddr}. However, these approaches primarily improve robustness or refine model outputs and do not explicitly model brush-like annotator behavior or progress-aware interaction. Interactive frameworks have used clicks or scribbles to iteratively refine segmentations \cite{wang2018interactive,sakinis2019deepgrow,diaz2023deepedit,monailabel2024}. Shahin \etal proposed an important interactive editing framework for intracardiac echocardiography segmentation, in which sparse user scribbles locally modify an existing segmentation while preserving regions away from the interaction \cite{shahinSparsePrecise2023}. This work demonstrated the value of sequential ultrasound editing, but its objective differs from ExiL: it focuses on local correction of model-generated masks using sparse prompts, whereas ExiL focuses on completing partial brush-based expert intent and adapting the model from expert-approved refinements across an annotation workflow. Larger collaborative pipelines decompose annotation into initialization, correction, and quality estimation \cite{benenson2019large}, while newer methods learn sequential correction policies over simulated trajectories \cite{zhu2025segagent}. Domain-adaptive approaches further tailor refinement to modality-specific noise \cite{marchesoniacland:hal-03835733}. Despite this progress, progress-aware completion of dense brush-based expert annotations remains underexplored. We therefore proposed ExiL, a progressive learning framework that integrates progress conditioning, behavioral modeling, and iterative expert-approved adaptation.

\paragraph{Limitations}:
Despite recent progress, key limitations remain. Most interactive methods rely on sparse prompts such as clicks or scribbles \cite{wang2018interactive,sakinis2019deepgrow,diaz2023deepedit,monailabel2024,shahinSparsePrecise2023}, which do not fully match real expert workflows that often use coarse brush-like or dense region-level annotations. Such approaches are valuable for local correction, yet they are not primarily designed to infer missing anatomical structure from incomplete brush masks or to personalize model behavior from expert-approved updates. In addition, many approaches model interaction as a sequence of corrections without explicitly representing evolving expert intent during annotation. Finally, learning expert-like policies would require intermediate annotation states and correction trajectories rather than only final masks, but such datasets are rarely available in medical imaging. As a result, existing methods depend on final labels or synthetic simulations that incompletely reflect authentic expert behavior. To address these limitations, this study proposes ExiL, a progress-aware adaptive refinement framework for US bone segmentation. \paragraph{Contributions}: Specifically, the framework introduces a realistic simulation strategy for generating partial and imperfect expert-like brush annotations, a mask-conditioned refinement network for boundary completion, online expert-approved adaptation, and a quantitative and qualitative feasibility evaluation showing gains over baseline refinement models and reduced manual annotation effort.

\section{Materials and Methods}
\subsection{Cadaveric Dataset and Ground-Truth Labels}
We used US bone annotations derived from the UltraBones100k dataset \cite{ultrabones100k}, 
which provides highly accurate bone surface labels generated via CT-to-US spatial alignment. 
This automated labeling strategy produces geometrically precise ground truth that is well suited for 
evaluating boundary-sensitive segmentation models. Although UltraBones100k contains over 100k images, our goal is to study completion learning  under realistic data scarcity conditions where expert annotation is scarce. Therefore, we intentionally restricted 
training to approximately 2.5k frames (sampled with stride 2 from ~5k images). This design reflects the 
clinical motivation of our work: if large-scale fully labeled datasets were readily available, 
expert-in-the-loop completion would be less necessary. All baseline models of this study, including nnU-Net and the SAM-style interactive model, were trained or fine-tuned 
using same limited subset to ensure fair comparison.

\subsection{Synthetic Dataset: Expert-imitating Annotation Synthesis}
Realistic modelling of imperfect annotations is critical for training a completion network that generalizes to expert behaviour. Prior interactive segmentation methods simulate clicks or scribbles to improve efficiency and robustness \cite{wang2018interactive,sakinis2019deepgrow,benenson2019large}, but they do not reflect the full brush-based masks commonly used in US labeling. To address the lack of intermediate annotation states, we developed a synthetic expert-like brush simulator that generates incomplete masks from clean ground-truth annotations. For each ground-truth mask, we first compute a signed distance field (SDF) to encode proximity to the bone boundary. A target coverage ratio is then sampled from predefined ranges (10--30\%, 40--60\%, and 70--90\%) to emulate both early rough painting and later refinement. Rather than using simple morphological corruption, we generate a smooth spatial bias field by applying Gaussian smoothing ($\sigma=5$ px) to random noise, creating uneven inward painting patterns resembling manual annotation. Brush-like masks are produced with stochastic stroke trajectories: starting points are sampled uniformly inside the ground-truth mask, paths evolve through temporally smoothed random motion with direction momentum of 0.2, step size of 2 px, and stroke thickness sampled uniformly from 3--5 px. Additional operations introduce missing regions, incomplete coverage, and limited boundary leakage. Unlike naive erosion/dilation or click-based simulation \cite{wong2024scribbleprompt}, this simulator models structured annotation via SDF-guided perturbations.
\begin{figure}[t]
\centering
\includegraphics[width=0.85\linewidth]{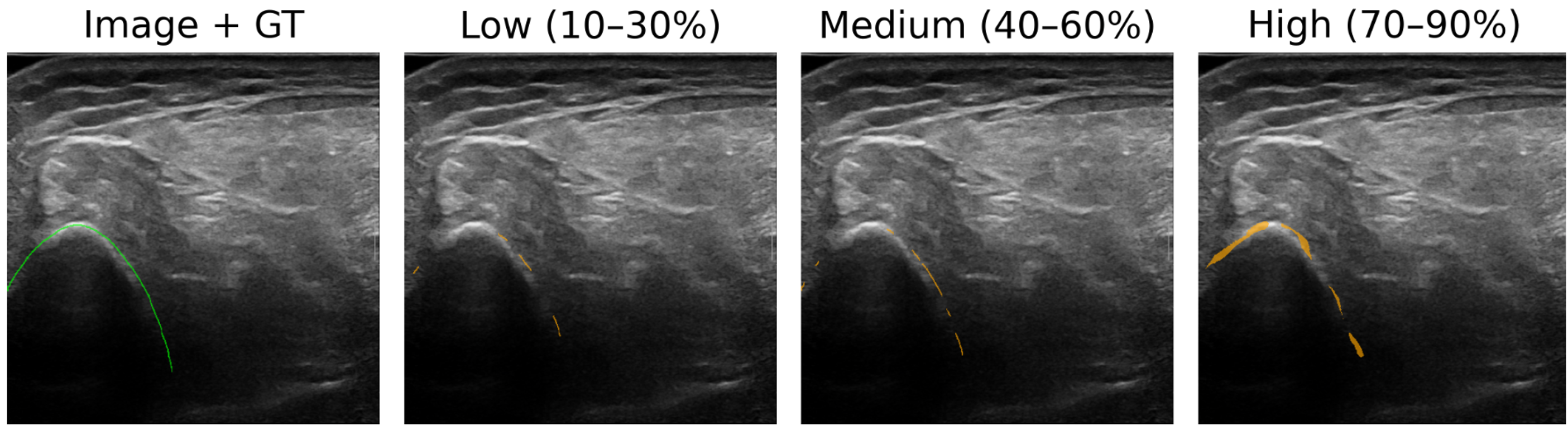}
\caption{Examples of synthetic expert-like brush annotations generated by the proposed simulator. Different coverage levels are sampled to mimic early-stage rough painting and later-stage refinement behaviours.}
\label{fig:brush_simulator}
\end{figure}

\subsection{nnUnet: Mask-Conditioned Annotation Completion Network}
The completion model is implemented as a U-Net architecture with approximately 7.8M trainable parameters. The input tensor was constructed by concatenating multiple channels depicted in Fig.\ref{fig:experiments}(a). The distance-to-brush channel provides geometric context indicating how far each pixel lies from the current annotation, while the edge map emphasizes high-frequency boundary cues important for aligning predictions with the bone interface. Unlike earlier variants of the framework, no explicit progress encoding is used; refinement behaviour is learned implicitly through trajectory-aware training. Encoder–decoder pathways with skip connections allow the network to integrate global anatomical context with fine-grained boundary information while preserving expert intent.
The network was trained for 25 epochs using a batch size of 4 on approximately 5,000 US image--mask pairs sampled at stride 2. Input images were resized to $512 \times 512$ pixels. Optimization was performed using the Adam optimizer with standard learning rate scheduling. Synthetic annotations generated by the brush simulator were used for initial training, followed by expert-in-the-loop updates during deployment.

\subsection{Expert-in-the-Loop (ExiL) Framework and Training}
\begin{figure}[t]
    \centering
    \includegraphics[
        width=\linewidth,
        trim=20 0 0 0,
        clip
    ]{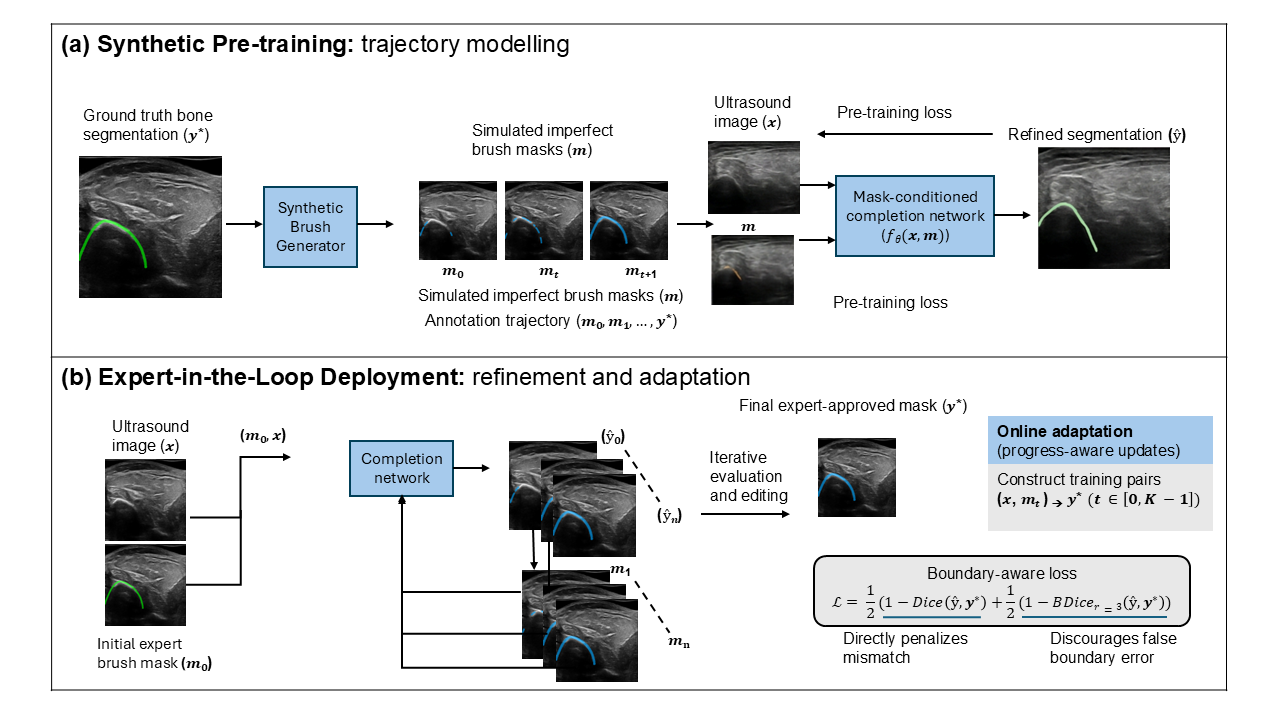}
    \caption{Pipeline overview of the proposed ExiL framework.}
    \label{fig:pipeline}
\end{figure}

We propose an Expert-in-the-Loop (ExiL) framework for US bone segmentation that completes and refines imperfect brush-based annotations through progressive expert interaction. Rather than treating rough masks as noisy labels, ExiL models them as structured intermediate states along an annotation trajectory. Given an US image $x$ and an expert brush mask $m$, a mask-conditioned completion network predicts a refined segmentation
$\hat{y} = f_{\theta}(x,m).$
During annotation, the expert iteratively edits the mask and re-invokes the model until an acceptable result is obtained. Once a final expert-approved mask $y^{*}$ is available, the model is updated in an \textit{expert mode} using only expert-edited input masks, preventing feedback from potentially corrupted model outputs. The encoder is frozen during deployment to preserve the generalizable representation learned during pretraining, while decoder weights are updated online to adapt to expert-specific refinements. The model is updated immediately after each expert-approved mask using five gradient steps with a learning rate of $10^{-4}$, matching the pretraining setting and limiting representation drift. The framework consists of two stages: (i) pretraining on synthetic expert-like brush simulations and (ii) deployment-time adaptation through trajectory-aware updates (Fig.~\ref{fig:pipeline}). During deployment, annotation begins from an initial mask $m_0$, followed by iterative predictions and expert corrections until convergence. For a trajectory $\{m_0,m_1,\ldots,m_{K-1}\}$, training pairs are formed as $(x,m_t) \rightarrow y^*, \quad t \in [0,K-1]$, so that each intermediate expert-edited mask learns to map toward the final approved segmentation. To further stabilize refinement, synthetic intermediate pairs $(x,m_a)\rightarrow m_b$ are also sampled, where $m_b$ is more refined than $m_a$, teaching the model the direction of correction and improving boundary convergence. To encourage accurate contours, ExiL uses a boundary-aware loss for both the base nnU-Net and ExiL fine-tuning:
\begin{equation}
\mathcal{L}
=
\frac{1}{2}\Big(1-\mathrm{Dice}(\hat{y},y^{*})\Big)
+
\frac{1}{2}\Big(1-\mathrm{BDice}_{r=3}(\hat{y},y^{*})\Big).
\label{eq:loss}
\end{equation}
Here, $\hat{y}$ is the predicted mask and $y^{*}$ is the ground-truth mask. This objective jointly enforces region overlap and contour alignment while tolerating negligible offsets ($\leq 3$ px). An additional penalty suppresses false positives above the ground-truth interface, reducing boundary leakage and improving anatomical consistency.

\subsection{Validation Studies}
\begin{figure}
    \centering
    \begin{tabular}{cc}
        \includegraphics[width=0.33\linewidth]{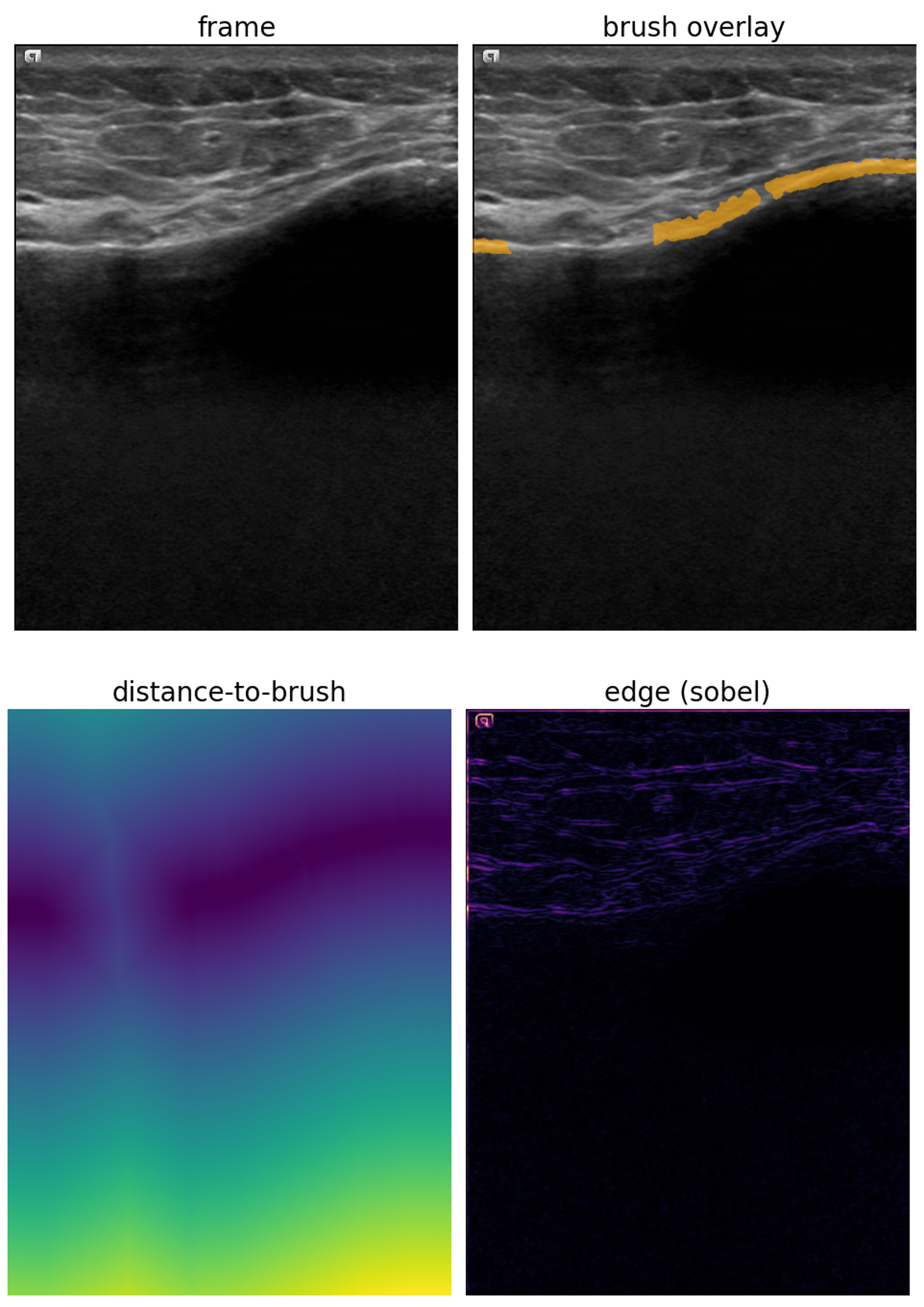} & \includegraphics[width=0.45\linewidth]{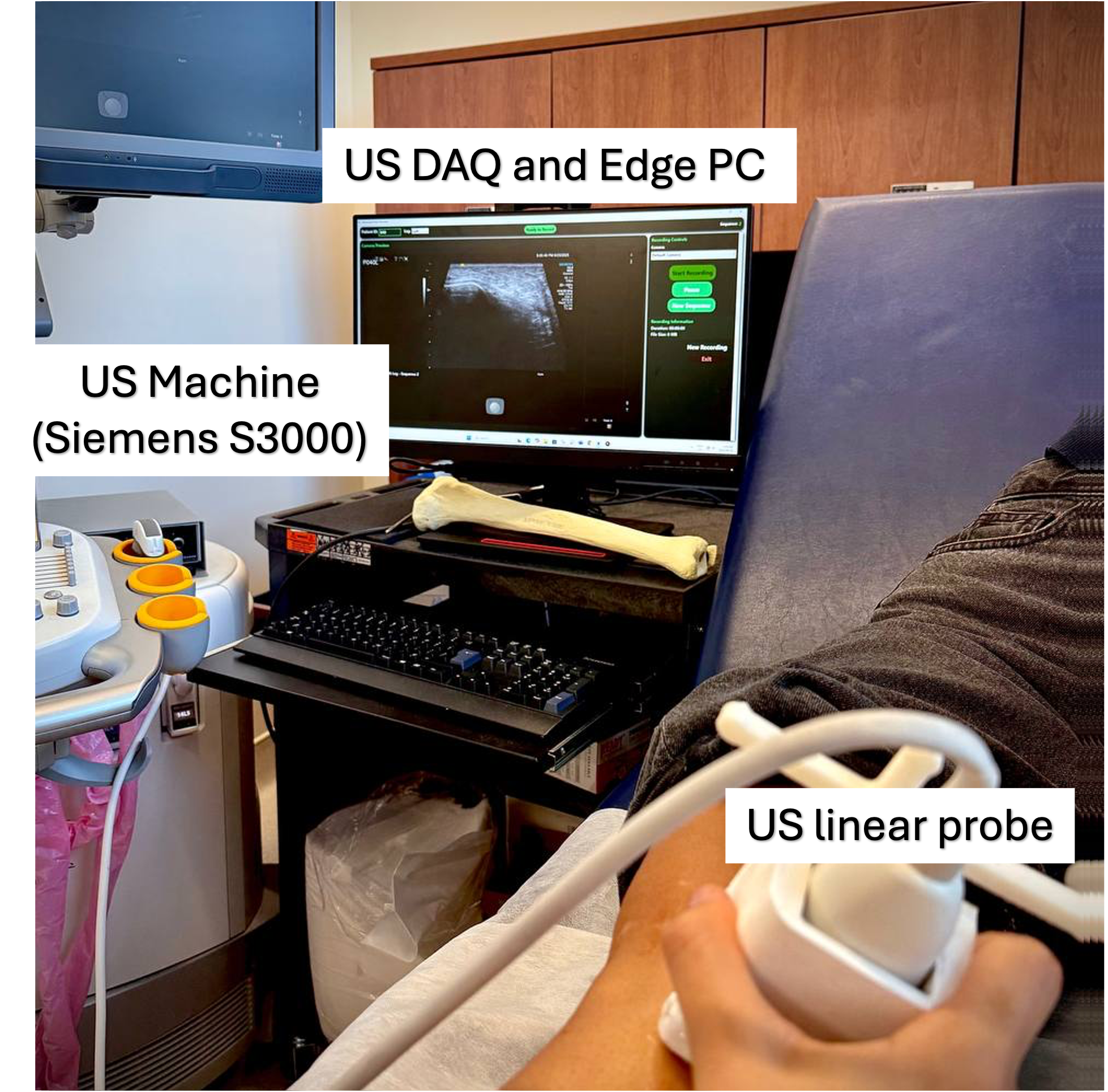} \\
         (a)&(b) 
    \end{tabular}
    
    \caption{(a) Visualization of input channels used by the mask-conditioned completion network and (b) the experimental setup used in this study for US data acquisition.}
    \label{fig:experiments}
\end{figure}

Two evaluation datasets were used for distinct purposes. Quantitative segmentation metrics, including Dice and boundary error, were evaluated using UltraBones100k cadaver data with ground-truth bone labels derived from CT-to-US alignment. Annotation efficiency was measured separately on an institutional ethics-approved prospective volunteer dataset (eRAP No. 25-05-109) consisting of three participants, with 6 US sequences per participant and 20 frames per sequence, for 360 annotated frames in total. These evaluation sequences were not used during training or model development. All images were resized to $512\times512$ and models were trained on up to 5{,}000 image--mask pairs sampled at stride 2.
Fig. \ref{fig:experiments}(b) depicts the experimental setup used in this study to acquire US images of the tibia.
To assess annotation efficiency under realistic conditions, a single expert annotator used all methods within the same interface, starting each frame from a rough brush mask and iteratively refining it until predefined acceptance criteria were met. The same stopping rules and visual guidelines were applied across methods, frames were randomized to reduce learning bias, and annotation time was measured from the first brush stroke to the final approved mask. Accordingly, the reported time reduction should be interpreted as a controlled within-annotator feasibility result rather than a multi-user generalization estimate. For quantitative comparison independent of expert edits, we also built a synthetic benchmark by generating imperfect brush masks at controlled coverage levels from ground truth and evaluating each model's \emph{autofill output} before correction. Under the same limited-data setting ($\approx$2.5k frames), we compared Manual-only, a fully automatic nnU-Net without rough-mask conditioning, a mask-conditioned Base Completion Model trained only on synthetic data, ExiL v1 with expert-approved updates, ExiL v2 with trajectory-aware learning, and a SAM-style interactive baseline restricted to a single refinement per frame, where synthetic rough masks were converted into sparse foreground prompts with prompt count proportional to coverage.

\section{Results and Discussion}

To evaluate completion quality independently of expert edits, we generated synthetic rough masks at three coverage levels: 10--30\%, 40--60\% and 70--90\%. Coverage is defined as the fraction of ground-truth pixels contained within the synthetic brush input. Each model received the same rough mask as input (when applicable), and we evaluated only the \emph{autofill prediction} prior to any manual correction. We report Dice and boundary error (in pixels), where lower boundary error indicates better alignment with the true bone contour. We then measured annotation efficiency on the prospective volunteer data using the standardized single-expert protocol described above. We report average time per frame and the number of refinement cycles (autofill invocations) required until acceptance.

\begin{table}[t]
\centering
\caption{Autofill performance on synthetic rough masks at different coverage levels.}
\label{tab:coverage}
\resizebox{\linewidth}{!}{
\begin{tabular}{lcccccccc}
\hline
& \multicolumn{2}{c}{10--30\%} & \multicolumn{2}{c}{40--60\%} & \multicolumn{2}{c}{70--90\%} & \multicolumn{2}{c}{Overall} \\
Method & Dice & BoundErr(px) & Dice & BoundErr(px) & Dice & BoundErr(px) & Dice & BoundErr(px) \\
\hline
nnU-Net (automatic) & 0.60 & 6.8 & 0.64 & 6.1 & 0.67 & 5.6 & 0.64 & 6.2 \\
SAM-style interactive & 0.66 & 6.2 & 0.72 & 5.4 & 0.76 & 4.8 & 0.71 & 5.5 \\
Base Completion & 0.74 & 5.1 & 0.80 & 4.2 & 0.84 & 3.6 & 0.79 & 4.3 \\
ExiL v1 (expert-updated) & 0.80 & 3.6 & 0.86 & 2.9 & 0.89 & 2.4 & 0.85 & 3.0 \\
ExiL v2 (trajectory-aware) & \textbf{0.83} & \textbf{3.2} & \textbf{0.88} & \textbf{2.6} & \textbf{0.90} & \textbf{2.2} & \textbf{0.87} & \textbf{2.7} \\
\hline
\end{tabular}}
\end{table}

\begin{figure}[t]
\centering
\begin{tabular}{cc}
\includegraphics[width=0.6\linewidth]{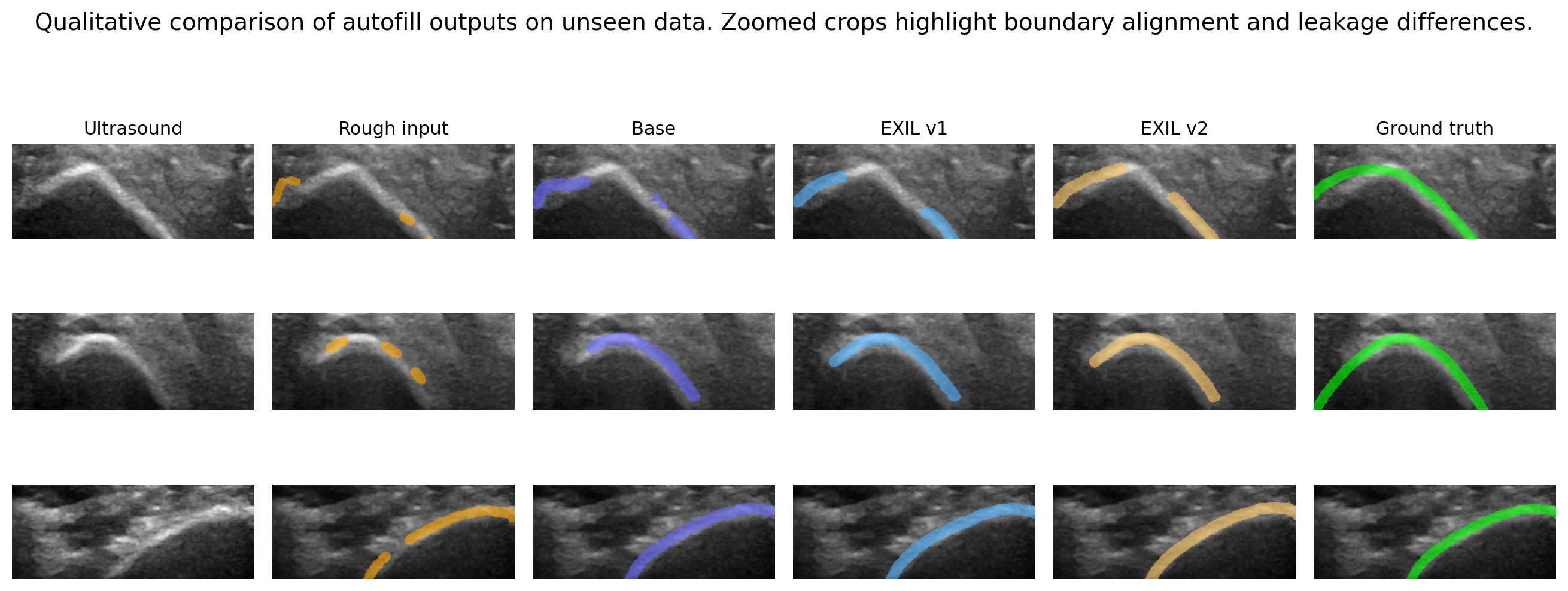}&
\includegraphics[width=0.35\linewidth]{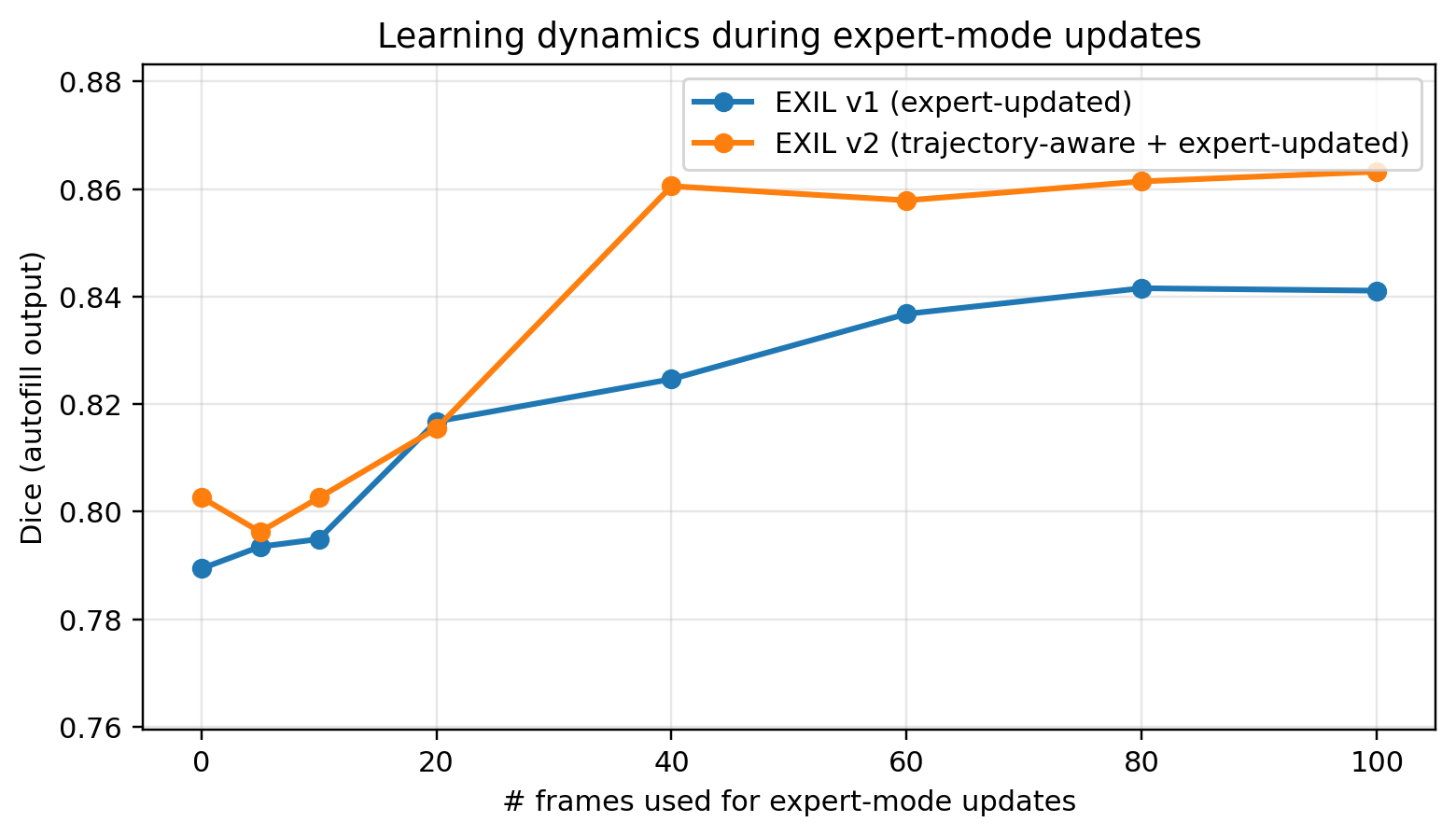}\\
(a)&(b)
\end{tabular}
\caption{(a) Model adaptation under expert-mode online updates across sequential frames, (b) Dice score improvement with increasing number of annotated frames.}
\label{fig:qualitative}
\end{figure}

\begin{table}[t]
\centering
\caption{Expert annotation efficiency on unseen data.}
\label{tab:time}
\footnotesize
\setlength{\tabcolsep}{3pt}
\resizebox{.7\columnwidth}{!}{%
\begin{tabular}{lccc}
\hline
Method & \shortstack{Avg Time /\\ Frame (s)} & \shortstack{Reduction\\ (\%)} & \shortstack{Autofill Cycles\\ / Frame} \\
\hline
Manual-only & 60.0 & -- & -- \\
SAM-style assisted & 38.0 & 36.7 & 2.9 \\
Base (no ExiL) & 41.0 & 31.7 & 3.4 \\
ExiL v1 (expert mode) & 28.0 & 53.3 & 2.2 \\
ExiL v2 (trajectory-aware) & \textbf{20.0} & \textbf{66.7} & \textbf{1.6} \\
\hline
\end{tabular}%
}
\end{table}
The expert-in-the-loop variants substantially reduced annotation time, with the trajectory-aware approach requiring fewer interaction cycles to reach an acceptable mask, consistent with improved completion quality in Table~\ref{tab:coverage}. To isolate contributions of key components, we ablated 
(i) boundary-aware losses, 
(ii) expert-mode updates, and 
(iii) trajectory-aware training. 
Metrics for Base, ExiL v1, and ExiL v2 correspond exactly to the values reported in Table~\ref{tab:ablation_combined}(a). Expert-mode updates produced the largest performance gains, confirming that learning from real expert interaction patterns is the primary driver of improvement. Boundary-aware losses reduced boundary misalignment, while trajectory aware supervision provided additional refinement stability, particularly in the low-coverage regime. We further evaluated the importance of geometric and edge cues by ablating input channels (Table \ref{tab:ablation_combined}(b)). Here, $I$ denotes the US image, $M$ the rough brush mask, $D(M)$ the distance-to-brush map, and $E$ the edge map. Distance-to-brush features improved completion under low coverage by providing geometric context, while edge cues improved boundary snapping. Using both cues achieved the best boundary alignment. Fig.~\ref{fig:qualitative}(a) illustrates representative examples from the held-out cadaver set. Each example shows the US image, the rough brush input, and the autofill outputs from the different models. Overall, ExiL v2 produced masks that were more consistently aligned with the upper bone interface. However, performance strongly depended on the prompt coverage. When the initial mask coverage was high, all models, including the base model, performed comparably well. In contrast, performance differences became more pronounced under low-coverage prompts, where ExiL v2 demonstrated improved boundary alignment and more complete mask. 
 \begin{table*}[t]
\centering
\caption{Ablation studies of training components and input channels.}
\label{tab:ablation_combined}

\begin{minipage}[t]{0.485\textwidth}
\centering
\small
\textbf{(a) Ablation of training components}\\[3pt]
\setlength{\tabcolsep}{4pt}
\resizebox{\linewidth}{!}{%
\begin{tabular}{lcccc}
\hline
& \multicolumn{2}{c}{10--30\% Coverage} & \multicolumn{2}{c}{Overall} \\
Variant & Dice & BoundErr(px) & Dice & BoundErr(px) \\
\hline
Base completion model      & 0.74 & 4.9 & 0.79 & 4.3 \\
+ Boundary-aware loss      & 0.76 & 4.4 & 0.81 & 3.8 \\
ExiL v1 (expert-updated)   & 0.80 & 3.6 & 0.85 & 3.0 \\
ExiL v2 (trajectory-aware) & \textbf{0.83} & \textbf{3.2} & \textbf{0.87} & \textbf{2.7} \\
\hline
\end{tabular}%
}
\end{minipage}
\hfill
\begin{minipage}[t]{0.485\textwidth}
\centering
\small
\textbf{(b) Input-channel ablation}\\[3pt]
\setlength{\tabcolsep}{4pt}
\resizebox{\linewidth}{!}{%
\begin{tabular}{lcc|cc}
\hline
& \multicolumn{2}{c|}{10--30\% Coverage} & \multicolumn{2}{c}{Overall} \\
Input & Dice & BoundErr(px) & Dice & BoundErr(px) \\
\hline
$I + M$                    & 0.70 & 5.7 & 0.77 & 4.9 \\
$I + M + D(M)$             & 0.73 & 5.1 & 0.79 & 4.5 \\
$I + M + E$                & 0.74 & 4.9 & 0.80 & 4.3 \\
$I + M + D(M) + E$ (full)  & \textbf{0.74} & \textbf{4.9} & \textbf{0.79} & \textbf{4.3} \\
\hline
\end{tabular}%
}
\end{minipage}
\end{table*}

As summarized in Table 1 and visualized in Fig.~\ref{fig:qualitative}, the model maintains strong performance under medium coverage, achieving results comparable to the high-coverage setting. This indicates that the trajectory-aware training strategy enables effective spatial inference even when supervision is partially available. Under low-coverage inputs, performance degradation becomes more apparent. In some cases, the model prioritizes advancing along the predicted trajectory rather than fully completing the missing mask regions. While this behavior reflects learned temporal consistency, it may lead to incomplete local refinement when spatial cues are highly sparse. Finally, we evaluated how performance evolved with sequential expert interaction. Fig.~\ref{fig:qualitative}(b) plots boundary error (and optionally Dice) of autofill outputs as a function of the number of annotated frames used for expert-mode updates. The model exhibited consistent improvement without degradation, suggesting stable adaptation to expert editing patterns. The results show that the proposed expert-in-the-loop framework markedly improves segmentation quality over rough expert annotations by learning structured correction from brush-based inputs and real expert edits rather than performing simple smoothing. In contrast to local point- or scribble-based correction systems, including Shahin \etal, that primarily edit existing masks around sparse prompts while preserving regions away from the interaction \cite{shahinSparsePrecise2023}, ExiL targets brush-mask-conditioned completion from partial expert intent and uses expert-approved online adaptation to improve behavior across sequential annotations. The largest gains occurred after expert-mode updates, underscoring the value of authentic interaction data, while ablations further showed that geometric distance encoding and boundary-aware losses improve boundary alignment, especially under low-coverage conditions. Although training was intentionally limited to a small subset to reflect realistic low-label settings, performance declined when initial masks were extremely sparse or anatomically inconsistent; future work will incorporate larger-scale real correction trajectories, extend to additional anatomical targets and modalities, and investigate temporal consistency for US sequences.

\section{Conclusion}

We presented an expert-in-the-loop autocorrection framework for US bone segmentation that learns to refine brush-based annotations through progressive expert interaction. By combining synthetic annotation simulation with expert-mode online updates, the proposed approach significantly improves both overlap and boundary accuracy while reducing structured annotation errors. The results demonstrate that modeling expert correction behavior provides a practical and scalable pathway for accelerating medical image annotation workflows and enhancing segmentation reliability in clinical US applications.

\section*{Disclosure of Interests}
The authors have no competing interests to declare that are relevant to the content of this article.

\bibliographystyle{unsrt}
\bibliography{Tavangar_etal_2026_MICCAI_refs}

\end{document}